\documentclass
[aps,prl,amsfonts,amssymb,twocolumn,amsmath,preprintnumbers,nofootinbib,floatfix,
showpacs]{revtex4-1}%
\usepackage[dvips]{graphics}
\usepackage{graphicx}
\usepackage{dcolumn}
\usepackage{bm}
\usepackage{amsmath}
\usepackage{amsfonts}
\usepackage{amssymb}
\usepackage{xcolor}
\usepackage{subfigure}
\usepackage{hyperref,hypcap}
\usepackage{braket}
\usepackage{commath}%
\providecommand{\U}[1]{\protect\rule{.1in}{.1in}}

\begin{document}

\title{Scaling parameters in anomalous and nonlinear Hall effects depend on temperature}
\author{Cong Xiao}
\affiliation{Department of Physics, The University of Texas at Austin, Austin, Texas 78712, USA}

\author{Hailong Zhou}
\affiliation{Department of Physics, The University of Texas at Austin, Austin, Texas 78712, USA}

\author{Qian Niu}
\affiliation{Department of Physics, The University of Texas at Austin, Austin, Texas 78712, USA}

\begin{abstract}
In the study of the anomalous Hall effect, the scaling relations between the
anomalous Hall and longitudinal resistivities play the central role. The
scaling parameters by definition are fixed as the scaling variable
(longitudinal resistivity) changes. Contrary to this paradigm, we unveil that
the electron-phonon scattering can result in apparent temperature-dependence
of scaling parameters when the longitudinal resistivity is tuned through
temperature. An experimental approach is proposed to observe this hitherto
unexpected temperature-dependence. We further show that this phenomenon also
exists in the nonlinear Hall effect in nonmagnetic inversion-breaking
materials and may help identify experimentally the presence of the side-jump
contribution besides the Berry-curvature dipole.
\end{abstract}
\maketitle

The anomalous Hall effect \cite{Nagaosa2010} has been a fruitful topic of
condensed matter physics, providing a paradigm widely-employed to understand
related nonequilibrium phenomena such as spin and valley Hall effects
\cite{Sinova2015,Xiao2007} and spin-orbit torque \cite{Xiao2017SOT-SBE}. In
time-reversal broken multiband electronic systems with strong spin-orbit
coupling, the anomalous Hall effect originates from both the momentum-space
Berry curvature and scattering off disorder
\cite{Sinitsyn2008,Sinitsyn2007,Kovalev2010}. In experiments the scaling
relations linking the anomalous Hall resistivity $\rho_{\text{AH}}$ to the
longitudinal resistivity $\rho$ play the central role in identifying various
contributions
\cite{Dheer1967,Coleman1974,Lee2004,Chun2007,Miyasato2007,Seemann2010,He2012,Tian2009,Pu2008,Hou2015,Yue2017,Li2016,Meng2016,Otani2019}%
.

The well-established theory taking into account a given type of weak-potential
static impurities \cite{KL1957,Sinitsyn2008,Nagaosa2010,Ado2017} results in
the scaling relation
\begin{equation}
-\rho_{\text{AH,0}}=\alpha_{0}\rho_{0}+\left(  c+c_{0}+c_{00}\right)  \rho
_{0}^{2}. \label{scaling-ei}%
\end{equation}
Henceforth the subscripts \textquotedblleft$0$\textquotedblright\ and
\textquotedblleft$1$\textquotedblright\ represent contributions from
electron-impurity and electron-phonon scattering, respectively. In this
scaling $\alpha_{0}$ arises from the skew scattering \cite{Smit}, $c$ is the
Berry-curvature contribution, and $c_{0}$ results from scattering-induced
coordinate-shift \cite{Sinitsyn2006}, namely the side-jump
\cite{Berger1970,Sinitsyn2007}. $c_{00}$ incorporates scattering-induced
contributions that are not related to coordinate-shift but share the same
scaling behavior as the side-jump one \cite{Sinitsyn2007,Ado2017}, and thereby
is referred to as the side-jump-like contribution \cite{note-sj-like}.
$\alpha_{0}$, $c$, $c_{0}$ and $c_{00}$ do not depend on the density of
scatterers thus serve as scaling parameters, and $\rho_{0}$ tuned via changing
the density of scatterers plays the role of a scaling variable.

On the other hand, in many experiments the resistivity is tuned through
temperature ($T$) in a wide range where the electron-phonon scattering is
important
\cite{Dheer1967,Coleman1974,Chun2007,Miyasato2007,Tian2009,Pu2008,Hou2015,Yue2017,Li2016}%
. For this scattering, most previous theoretical and experimental researches
suggest the scaling relation
\cite{Berger1970,Bruno2001,Leribaux1966,Hou2015,Dheer1967,Coleman1974}%
\begin{equation}
-\rho_{\text{AH,1}}=\left(  c+c_{1}+c_{11}\right)  \rho_{1}^{2},
\label{scaling-ep}%
\end{equation}
where the scaling parameters $c_{1}$ and $c_{11}$ are thought, according to
the aforementioned characteristic of side-jump and side-jump-like
contributions, to be independent of the density of phonons and thus of $T$
\cite{Berger1970,Bruno2001,Hou2015}.

\begin{figure}[tbh]
\includegraphics[width=0.8\columnwidth]{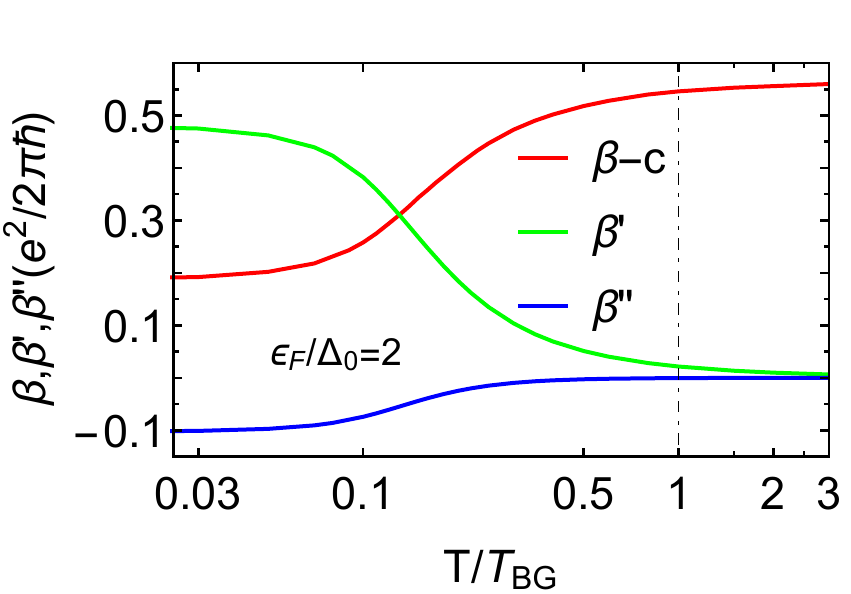}\caption{{}%
Temperature-dependence of $\beta$'s of Eq. (\ref{central-2}) in the 2D massive
Dirac model (\ref{model}) in the presence of both zero-range scalar impurities
and acoustic phonons. $T_{BG}$ is the Bloch-Gruneisen temperature
\cite{Kim2010} which plays the basic role instead of the Debye temperature in
2D metallic systems and marks qualitatively the lower boundary of the high-$T$
classical equipartition regime ($T>T_{BG}$) where the resistivity is linear in
$T$. When $T$ downs below the equipartition regime the $\beta$'s become $T$
dependent.}%
\label{fig:beta}%
\end{figure}

In the presence of both impurities and phonons, when assuming the
Matthiessen's rule $\rho=\sum_{i=0,1}\rho_{i}$, a two-variable scaling based
on the above two scalings reads \cite{Hou2015} ($\rho\gg\rho_{\text{AH}}$,
$\sigma_{\text{AH}}\simeq-\rho_{\text{AH}}/\rho^{2}$)
\begin{subequations}
\begin{equation}
\sigma_{\text{AH}}=\alpha_{0}\rho_{0}/\rho^{2}+c+\sum_{i=0,1}c_{i}\rho
_{i}/\rho+\sum_{i,j=0,1}c_{ij}\rho_{i}\rho_{j}/\rho^{2}. \label{scaling-HJYN}%
\end{equation}
Here $c_{10}+c_{01}$ represents the combined effect of scatterings off
impurities and phonons, and is also regarded to remain constant as $T$ changes
in previous studies \cite{Hou2015,Yue2017,Li2016,Meng2016,Otani2019}.

In this Letter we uncover that the above widely-accepted paradigm misses the
physics that $c_{1}$, $c_{11}$ and $c_{10}+c_{01}$ can be strongly $T$
dependent as $T$ drops below the high-$T$ classical equipartition regime where
$\rho_{1}\propto T$ \cite{Ziman1972}. In the minimal model of the anomalous
Hall effect, namely the two-dimensional (2D) massive Dirac model, Eq.
(\ref{scaling-HJYN}) is demonstrated with all $c$'s given explicitly (detailed
later) and reorganized into%
\begin{equation}
\sigma_{\text{AH}}-\alpha_{0}\sigma_{xx}^{2}/\sigma_{0}=\beta+\beta^{\prime
}\sigma_{xx}/\sigma_{0}+\beta^{\prime\prime}\left(  \sigma_{xx}/\sigma
_{0}\right)  ^{2},\label{central-2}%
\end{equation}
where $\sigma_{0}^{-1}=\rho_{0}$. The $T$ dependence of
\end{subequations}
\begin{align}
\beta\left(  T\right)   &  =c+c_{1}\left(  T\right)  +c_{11}\left(  T\right)
,\nonumber\\
\beta^{\prime}\left(  T\right)   &  =c_{0}-c_{1}\left(  T\right)
+c_{01}\left(  T\right)  +c_{10}\left(  T\right)  -2c_{11}\left(  T\right)
,\nonumber\\
\beta^{\prime\prime}\left(  T\right)   &  =c_{00}+c_{11}\left(  T\right)
-c_{01}\left(  T\right)  -c_{10}\left(  T\right)  ,
\end{align}
are shown in Fig. \ref{fig:beta}, although they are believed to be $T$
independent in the conventional paradigm of the anomalous Hall effect. Despite
that the specific $T$ dependent forms of $\beta$'s depend on fine details of
the model, the revealed possibility of the $T$ dependence of $c_{1}$, $c_{11}$
and $c_{10}+c_{01}$ is ubiquitous, as shown in the Supplemental Material
\cite{Supp}. 

This finding indicates that, the conventionally identified \textquotedblleft
scaling parameters\textquotedblright\ in Eqs. (\ref{scaling-ep}) --
(\ref{central-2}) are in fact allowed to vary with temperature when fitting
data. From a viewpoint of basic understanding, scaling parameters by
definition ought to be fixed as the scaling variable changes, thereby Eqs.
(\ref{scaling-ep}) -- (\ref{central-2}) can not be termed as \textquotedblleft
scaling relations\textquotedblright\ when $\rho$ is tuned through temperature.
On the practical side, since $\beta$, $\beta^{\prime}$ and $\beta
^{\prime\prime}$ do not depend on the density of impurities $n_{i}$, and thus
still serve as scaling parameters when $\rho$ is tuned by changing the density
of impurities, we propose an experimental procedure to observe the $T$
dependence of $\beta$'s.

Now we begin the concrete analysis of the 2D massive Dirac model
\cite{Sinitsyn2007}
\begin{equation}
\hat{H}_{0}=v\left(  \hat{\sigma}_{x}k_{x}+\hat{\sigma}_{y}k_{y}\right)
+\Delta_{0}\hat{\sigma}_{z}, \label{model}%
\end{equation}
where $\hat{\sigma}_{x,y,z}$ are the Pauli matrices, $\mathbf{k}=\left(
k_{x},k_{y}\right)  $ is the wave-vector, $v>0$ and $\Delta_{0}>0$ are model
parameters. We consider the electron-doped case with Fermi energy
$\epsilon_{F}>\Delta_{0}$. For the purpose of revealing the fact that the
$\beta$'s depend on $T$, the quasi-static treatment for acoustic phonons is
adequate. This approximation, in which the electron-phonon scattering is
treated as an elastic process, produces the correct low-$T$ Bloch-Gruneisen
law $\rho_{1}\sim T^{5}$ for three-dimensional (3D) metals
\cite{Ziman1972,Ashcroft,Marder}. When applied to the side-jump anomalous Hall
effect, the high-$T$ and low-$T$ asymptotic behaviors derived within this
approximation are the same as those obtained without this approximation
\cite{Xiao2019PhononSJ}. Quantitative deviations only appear\ in the
intermediate regime and are not essentially important.

To proceed, we employ the Boltzmann transport theory involving not only
on-shell (on the Fermi surface) but also off-shell (away from the Fermi
surface) Bloch states \cite{KL1957,Sinitsyn2007,Sinitsyn2008,Nagaosa2010}. In
the presence of scalar quasi-static disorder, the side-jump (sj) and
side-jump-like (sjl) contributions to the anomalous Hall conductivity of model
(\ref{model}) are obtained as \cite{Supp}%
\begin{equation}
\sigma_{\text{AH}}^{\text{sj}}=\frac{e^{2}}{4\pi\hbar}\sin^{2}\theta_{F}%
\cos\theta_{F}\left(  \tau_{0}^{-1}-\tau_{1}^{-1}\right)  \tau_{tr} \label{sj}%
\end{equation}
and
\begin{align}
\sigma_{\text{AH}}^{\text{sjl}}  &  =\frac{e^{2}}{64\pi\hbar}\sin^{4}%
\theta_{F}\cos\theta_{F}\left(  \tau_{0}^{-1}-\tau_{2}^{-1}\right)
\label{sj-like}\\
&  \times\left(  3\tau_{0}^{-1}-4\tau_{1}^{-1}+\tau_{2}^{-1}\right)  \tau
_{tr}^{2},\nonumber
\end{align}
respectively. Here $\cos\theta_{F}=\Delta_{0}/\epsilon_{F}$, $\sin\theta
_{F}=vk_{F}/\epsilon_{F}$. $\tau_{tr}^{-1}$ is the value of the inverse
transport relaxation time
\begin{equation}
\tau_{tr}^{-1}\left(  k\right)  =\frac{D_{k}}{\hbar}\int d\phi_{\mathbf{k}%
^{\prime}\mathbf{k}}\left\vert \langle u_{\mathbf{k}^{\prime}}|u_{\mathbf{k}%
}\rangle\right\vert ^{2}W_{\mathbf{k}^{\prime}\mathbf{k}}\left(  1-\cos
\phi_{\mathbf{k}^{\prime}\mathbf{k}}\right)  \label{tau-transport}%
\end{equation}
on the Fermi surface, and
\begin{equation}
\tau_{n}^{-1}\left(  k\right)  =\frac{D_{k}}{\hbar}\int d\phi_{\mathbf{k}%
^{\prime}\mathbf{k}}W_{\mathbf{k}^{\prime}\mathbf{k}}\cos\left(
n\phi_{\mathbf{k}^{\prime}\mathbf{k}}\right)  ,\text{ \ }n=0,1,2...,
\label{tau-i}%
\end{equation}
where $D_{k}$ is the density of states, $\phi_{\mathbf{k}^{\prime}\mathbf{k}}$
is angle between $\mathbf{k}$ and $\mathbf{k}^{\prime}$, $|u_{\mathbf{k}%
}\rangle$ is the spinor eigenstate in the positive band, and $W_{\mathbf{k}%
^{\prime}\mathbf{k}}$ is the plane-wave part of the lowest-Born-order
scattering rate $\left\vert \langle u_{\mathbf{k}^{\prime}}|u_{\mathbf{k}%
}\rangle\right\vert ^{2}W_{\mathbf{k}^{\prime}\mathbf{k}}$.

For quasi-static electron-phonon scattering one has $W_{\mathbf{k}^{\prime
}\mathbf{k}}^{\text{ep}}=\frac{2N_{\mathbf{q}}}{\text{V}}\left\vert
U_{\mathbf{k}^{\prime}\mathbf{k}}^{o}\right\vert ^{2}$, where $U_{\mathbf{k}%
^{\prime}\mathbf{k}}^{o}$ is the plane-wave part of the electron-phonon matrix
element, V is the volume (area in 2D) of the system, $N_{\mathbf{q}}$ is the
Bose occupation function for the phonon model with wave-vector $\mathbf{q}$
and energy $\hbar\omega_{q}$, and the factor $2$ accounts for the absorption
and emission of phonons \cite{Ziman1972}. To simplify the analysis we neglect
the Umklapp process, thus $\mathbf{q}=\mathbf{k}^{\prime}-\mathbf{k}$. In
comparison, for static impurities $W_{\mathbf{k}^{\prime}\mathbf{k}%
}^{\text{ei}}=n_{i}\left\vert V_{\mathbf{k}^{\prime}\mathbf{k}}^{o}\right\vert
^{2}$, with $V_{\mathbf{k}^{\prime}\mathbf{k}}^{o}$ the plane-wave part of the
matrix element of the impurity potential. Hereafter the superscript
\textquotedblleft ei/ep\textquotedblright\ means that the quantity is
contributed by the electron-impurity/-phonon scattering alone.

To obtain analytic results, we assume zero-range scalar impurities
($\left\vert V_{\mathbf{k}^{\prime}\mathbf{k}}^{o}\right\vert ^{2}=V_{i}^{2}$
is a constant), isotropic Debye phonons, and the deformation-potential
coupling for which a so-called electron-phonon coupling constant $\lambda
^{2}=2$V$^{-1}\left\vert U_{\mathbf{k}^{\prime}\mathbf{k}}^{o}\right\vert
^{2}/\hbar\omega_{q}$ exists \cite{Abrikosov,Rammer}. Then
\begin{equation}
W_{\mathbf{k}^{\prime}\mathbf{k}}=W_{\mathbf{k}^{\prime}\mathbf{k}}%
^{\text{ei}}+W_{\mathbf{k}^{\prime}\mathbf{k}}^{\text{ep}}=n_{i}V_{i}%
^{2}+\lambda^{2}k_{B}T\frac{z}{e^{z}-1}, \label{W-ep+ei}%
\end{equation}
where $z=\hbar\omega_{q}/k_{B}T=\frac{q}{2k_{F}}\frac{T_{BG}}{T}$, and
$T_{BG}=\hbar c_{s}2k_{F}/k_{B}$ ($c_{s}$ is the sound velocity) is the
Bloch-Gruneisen temperature \cite{Kim2010}. Model results of $\sigma
_{\text{AH}}^{\text{SJ}}\equiv\sigma_{\text{AH}}^{\text{sj}}+\sigma
_{\text{AH}}^{\text{sjl}}$ are obtained \cite{Supp} according to%
\begin{equation}
\tau_{n}^{-1}\tau_{tr}=\frac{\left(  \tau_{n}^{\text{ei}}\right)
^{-1}+\left(  \tau_{n}^{\text{ep}}\right)  ^{-1}}{\left(  \tau_{tr}%
^{\text{ep}}\right)  ^{-1}+\left(  \tau_{tr}^{\text{ep}}\right)  ^{-1}%
},\text{\ \ \ }n=0,1,2...
\end{equation}

Next we take into account the skew scattering (sk) from the third-order
non-Gaussian impurity correlator $n_{i}V_{i}^{3}$ of zero-range scalar
impurities \cite{Sinitsyn2007,Yang2011}:
\begin{equation}
\sigma_{\text{AH}}^{\text{sk}}=\frac{e^{2}}{16\pi\hbar}\sin^{4}\theta
_{F}\left(  \frac{\tau_{tr}}{\tau_{0}^{\text{ei}}}\right)  ^{2}D_{F}V_{i}%
\frac{\Delta_{0}\tau_{0}^{\text{ei}}}{\hbar}, \label{sk}%
\end{equation}
where $D_{F}$ is the density of states on the Fermi surface.

\begin{figure}[tbh]
\includegraphics[width=0.72\columnwidth]{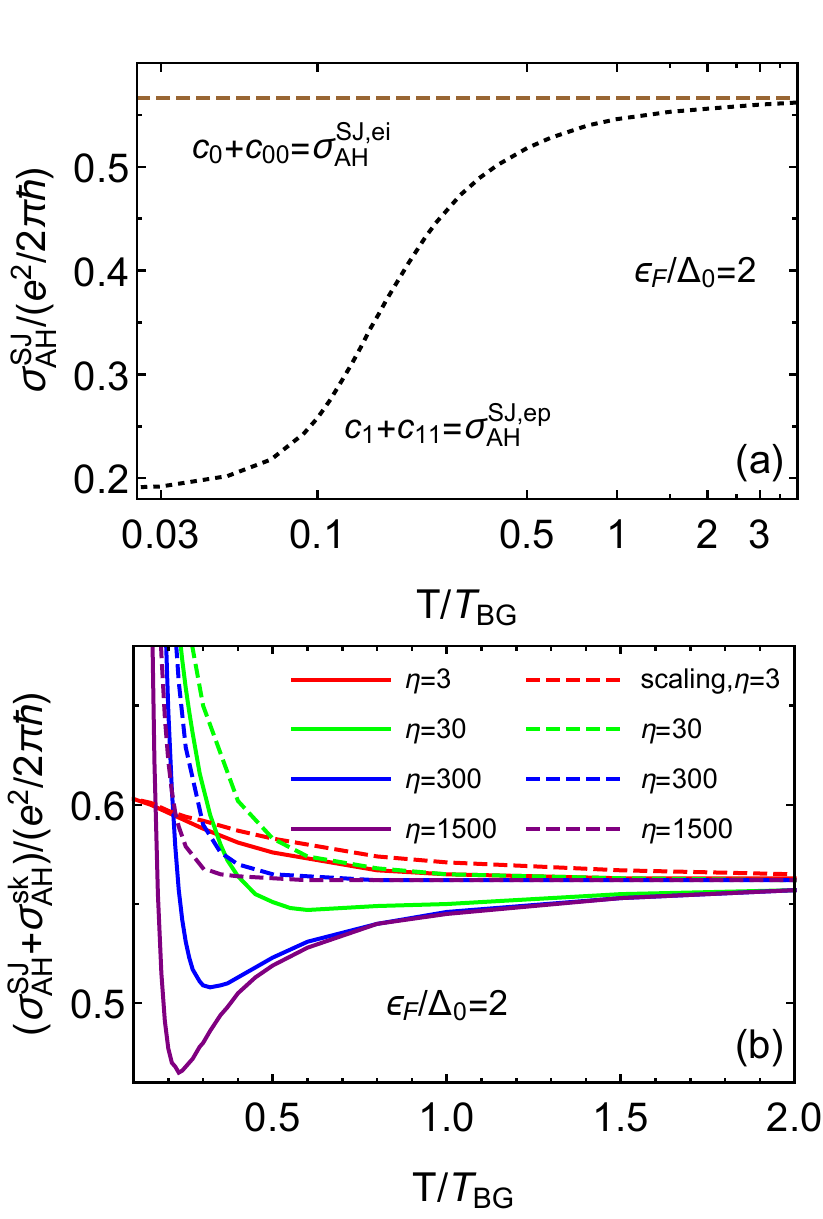}\caption{{}%
Temperature-dependence of (a) $\sigma_{\text{AH}}^{\text{SJ,ei}}$ and
$\sigma_{\text{AH}}^{\text{SJ,ep}}$, and of (b) $\sigma_{\text{AH}}%
^{\text{SJ}}+\sigma_{\text{AH}}^{\text{sk}}=\sigma_{\text{AH}}-c$ for
$\epsilon_{F}/\Delta_{0}=2$ in model (\ref{model}) in the presence of
zero-range scalar impurities and acoustic phonons. Larger values of the
parameter $\eta$ correspond to smaller impurity density. The dashed curves in
(b) are obtained by assuming the scaling relation (\ref{scaling-TYJ}) in the
presence of both scattering sources. In the weak scattering regime
$D_{F}\left\vert V_{i}\right\vert \ll1$ and $\Delta_{0}\tau_{0}^{\text{ei}%
}/\hbar\gg1$. Thus in the calculation of $\sigma_{\text{AH}}^{\text{sk}}$ we
take $D_{F}V_{i}=10^{-3}$, $\Delta_{0}\tau_{0}^{\text{ei}}/\hbar=10^{j+2}$ for
$\eta=3\times10^{j}$ ($j=0,1,2$), and $\Delta_{0}\tau_{0}^{\text{ei}}%
/\hbar=5\times10^{4}$ for $\eta=1500$ ($n_{i}$ is tuned).}%
\label{fig:T-dependence-sj+sk}%
\end{figure}

The model result thus takes the form of%
\begin{equation}
\sigma_{\text{AH}}-c-\sigma_{\text{AH}}^{\text{sk}}=\sum_{n}a_{n}^{\text{sj}%
}\tau_{n}^{-1}\tau_{tr}+\sum_{nn^{\prime}}b_{nn^{\prime}}^{\text{sjl}}\tau
_{n}^{-1}\tau_{n^{\prime}}^{-1}\tau_{tr}^{2},
\end{equation}
where $a_{n}^{\text{sj}}$ and $b_{nn^{\prime}}^{\text{sjl}}$ are readable from
Eqs. (\ref{sj}) and (\ref{sj-like}). By noting that $\rho_{0\left(  1\right)
}/\rho=\tau_{tr}/\tau_{tr}^{\text{ei(ep)}}$ and $\sigma_{\text{AH}}%
^{\text{sk}}=\alpha_{0}\rho_{0}/\rho^{2}$, the above equation can be cast into
Eq. (\ref{scaling-HJYN}), where the impurity-determined coefficients%
\begin{equation}
c_{0}=\sigma_{\text{AH}}^{\text{sj,ei}}=\sum_{n}a_{n}^{\text{sj}}\frac
{\tau_{tr}^{\text{ei}}}{\tau_{n}^{\text{ei}}},\text{ \ }c_{00}=\sigma
_{\text{AH}}^{\text{sjl,ei}}=\sum_{nn^{\prime}}b_{nn^{\prime}}^{\text{sjl}%
}\frac{\left(  \tau_{tr}^{\text{ei}}\right)  ^{2}}{\tau_{n}^{\text{ei}}%
\tau_{n^{\prime}}^{\text{ei}}},
\end{equation}
and $\alpha_{0}\sim D_{F}V_{i}\tau_{tr}^{\text{ei}}/\tau_{0}^{\text{ei}}$ are
independent of the density of impurities, while the phonon-determined
coefficients, namely%
\begin{equation}
c_{1}=\sigma_{\text{AH}}^{\text{sj,ep}}=\sum_{n}a_{n}^{\text{sj}}\frac
{\tau_{tr}^{\text{ep}}}{\tau_{n}^{\text{ep}}},\text{ \ }c_{11}=\sigma
_{\text{AH}}^{\text{sjl,ep}}=\sum_{nn^{\prime}}b_{nn^{\prime}}^{\text{sjl}%
}\frac{\left(  \tau_{tr}^{\text{ep}}\right)  ^{2}}{\tau_{n}^{\text{ep}}%
\tau_{n^{\prime}}^{\text{ep}}}, \label{c-phonon}%
\end{equation}
are $T$ dependent at low temperatures below the high-$T$ equipartition regime
as shown in Fig. \ref{fig:T-dependence-sj+sk}(a). This implies that Eq.
(\ref{scaling-ep}) can not be theoretically viewed as a scaling relation,
since the conventionally identified \textquotedblleft scaling
paramter\textquotedblright\ $c_{1}+c_{11}$ in fact changes as $\rho_{1}$
varies with temperature. Equations (\ref{scaling-HJYN}) and (\ref{central-2})
suffer from the same situation. Meanwhile, the combined contribution from the
impurity and phonon scatterings to the side-jump-like anomalous Hall
conductivity%
\begin{equation}
c_{01}+c_{10}=\sum_{nn^{\prime}}b_{nn^{\prime}}^{\text{sjl}}\left(  \frac
{\tau_{tr}^{\text{ei}}}{\tau_{n}^{\text{ei}}}\frac{\tau_{tr}^{\text{ep}}}%
{\tau_{n^{\prime}}^{\text{ep}}}+\frac{\tau_{tr}^{\text{ep}}}{\tau
_{n}^{\text{ep}}}\frac{\tau_{tr}^{\text{ei}}}{\tau_{n^{\prime}}^{\text{ei}}%
}\right)
\end{equation}
also depends on $T$ below the equipartition regime. The $T$ dependence of
$c_{1}$, $c_{11}$\ and $c_{01}+c_{10}$\ yields that of the $\beta$'s shown in
Fig. \ref{fig:beta}.

The qualitative picture for the appearance of the $T$ dependence is simple:
the side-jump and side-jump-like contributions are conventionally viewed as
independent of scattering time for a given source of scattering
\cite{Freimuth2011,Kovalev2010,Berger1970}, but in general they are just
zeroth-order homogeneous terms of scattering time \cite{Yang2011}, as is
apparent in Eqs. (\ref{sj}) and (\ref{sj-like}). In the equipartition regime
the $T$ dependence of electron-phonon scattering times in the denominator and
numerator of these zeroth-order homogeneous terms are the same ($T^{-1}$), and
thereby drop out of $\beta$'s. While at lower temperatures below the
equipartition regime the bosonic nature of phonon occupation number makes the
$T$ dependence irreducible even in the zeroth-order homogeneous terms.

The model analysis also offers a perspective to understand why the
conventional idea of $T$ independent scaling parameters works practically in
tuning-$T$ experiments. Because in the high-$T$ regime $W^{\text{ep}}%
=\lambda^{2}k_{B}T$ drops out of $\tau_{tr}/\tau_{n}$, the value of
$\sigma_{\text{AH}}^{\text{SJ}}$ contributed by phonons coincides with that
contributed by zero-range scalar impurities. This value also applies in the
presence of both scattering sources since $W=n_{i}V_{i}^{2}+\lambda^{2}k_{B}T$
for this case also drops out of $\tau_{tr}/\tau_{n}$. Therefore, in the
high-$T$ regime the scalar zero-range impurities and acoustic phonons are
indistinguishable in inducing $\sigma_{\text{AH}}^{\text{SJ}}$, namely,
$c_{1}=c_{0}$ and $c_{ij}=c_{00}$. Then, if the electron-phonon scattering
related $c$'s took $T$ independent values, just following the conventional
idea, the scaling relation would hold and read%
\begin{equation}
\sigma_{\text{AH}}=\alpha_{0}\sigma_{0}^{-1}\sigma_{xx}^{2}+c+c_{0}+c_{00}.
\label{scaling-TYJ}%
\end{equation}
As is shown by the dashed curves in Fig. \ref{fig:T-dependence-sj+sk}(b), this
scaling relation well describes the $T$ dependence of the anomalous Hall
conductivity in moderately dirty systems with smaller $\eta$. Here the
dimensionless parameter $\eta=\lambda^{2}k_{B}T_{BG}/n_{i}V_{i}^{2}$ is
introduced to denote the purity of system: larger $\eta$ means smaller $n_{i}%
$. In our model case $T_{BG}\ $is the lower boundary $T_{L}$ of the
equipartition regime where $\rho_{1}\propto T$, thus $\eta\simeq\tau
_{0}^{\text{ei}}/\tau_{0}^{\text{ep}}\left(  T_{L}\right)  =\tau
_{tr}^{\text{ei}}/\tau_{tr}^{\text{ep}}\left(  T_{L}\right)  $, taking the
instructive form of
\begin{equation}
\eta\simeq\frac{\rho_{1}\left(  T_{L}\right)  }{\rho_{0}}=\frac{\rho\left(
T_{L}\right)  -\rho_{0}}{\rho_{0}},
\end{equation}
which can be read out conveniently from transport experiments. Given the
smaller $\eta$ (usually $<2$) of the samples used in previous tuning-$T$
experiments, the above argument provides a clue to understand the approximate
validity of scaling relations in these experiments.

As is shown by Fig. \ref{fig:T-dependence-sj+sk}(b), the deviation of
$\sigma_{\text{AH}}$ caused by assuming the scaling relation is apparent only
in high-purity systems. By contrast, it is worthwhile to emphasize that in Eq.
(\ref{central-2}) the $\beta$'s do \textit{not} depend on $\rho_{0}$, thus
their $T$ dependence is anticipated to show up irrespective of the sample quality.

Next we propose an experimental procedure to observe the predicted $T$
dependence of $\beta$'s, based on the recently developed thin film approach in
the study of the anomalous Hall effect \cite{Yue2017,Tian2009,Hou2015}. In
this approach the effective impurity density can be continuously manipulated
by tuning the thickness of single crystalline magnetic thin films (the Curie
temperature is assumed to be much higher than the Debye temperature, which is
the case of Fe and Co), meanwhile the electronic band structure does not
change in the thickness range. In the low-$T$ limit $\sigma_{xx}=\sigma_{0}$,
and Eq. (\ref{central-2}) reduces to the linear scaling $\sigma_{\text{AH}%
}=\alpha_{0}\sigma_{0}+c+c_{0}+c_{00}$, thereby $\alpha_{0}$ can be extracted
by tuning $\sigma_{0}$ through the film thickness. Because in Eq.
(\ref{central-2}) the $\beta$'s\ are still scaling parameters that remain
unchanged\ \textit{when the film thickness is tuned}, it is reasonable to plot
$(\sigma_{\text{AH}}-\alpha_{0}\sigma_{0}^{-1}\sigma_{xx}^{2})$ versus
$\sigma_{0}^{-1}\sigma_{xx}$ through tuning the film thickness for every
chosen fixed temperature. One can then extract the $\beta$'s for different
temperatures from the high-$T$ equipartition regime $T>T_{L}$ (experiments in
common 3D metals often show $T_{L}\simeq T_{D}/3$ as the lower boundary of the
$\rho_{1}\propto T$ regime, with $T_{D}$ the Debye temperature
\cite{Ziman1960,White1958}) down to the low-$T$ residual-resistivity regime.
The $T$-variation curves of $\beta$'s are thus obtained. The predicted $T$
independent values of $\beta$'s at $T>T_{L}$ can be determined first, whereas
their $T$-dependence can be observed as temperature downs below $T_{L}$.

Finally we extend the discussion to the nonlinear Hall effect -- a
second-order Hall current response to the electric field $E_{x}$ in
nonmagnetic systems with inversion breaking
\cite{Fu2015,Low2015,Lu2018,Yan2018,Low2018,Ma2019,Mak2019,Facio2018,Xu2018,Zhou2019}%
: $j_{y}=\chi_{yxx}E_{x}E_{x}$, with $\chi_{yxx}$ the response coefficient.
The systematic Boltzmann analysis of $\chi_{yxx}$, which is of the linear
order of scattering time, incorporates the Berry-curvature dipole (bcd)
mechanism \cite{Fu2015} and the nonlinear generalizations of the side-jump,
side-jump-like and skew scattering contributions
\cite{Du2018,Konig2018,Fu2018,Deyo2009,Sodemann2019,Xiao2019NLHE}. Naturally,
equation (\ref{scaling-HJYN}) has been extended to the nonlinear Hall effect
in the dc limit as \cite{Du2018}
\begin{equation}
\frac{V_{y}^{N}}{(V_{x}^{L})^{2}}=C+A_{0}\frac{\rho_{0}}{\rho_{xx}^{2}}%
+\sum_{i=0,1}C_{i}\frac{\rho_{i}}{\rho_{xx}}+\sum_{i,j=0,1}C_{ij}\frac
{\rho_{i}\rho_{j}}{\rho_{xx}^{2}},
\end{equation}
where $V_{y}^{N}$ and $V_{x}^{L}$ are the nonlinear Hall and linear
longitudinal voltage, respectively, and $V_{y}^{N}/\left(  V_{x}^{L}\right)
^{2}=\chi_{yxx}\rho_{xx}$. Here the notation $\rho_{xx}$ is used instead of
$\rho$, considering the low-symmetry of the materials for observing the
nonlinear Hall effect \cite{Ma2019,Mak2019}. Equivalently,
\begin{equation}
V_{y}^{N}/(V_{x}^{L})^{2}-A_{0}\sigma_{xx}^{2}/\sigma_{0}=B+B^{\prime}%
\sigma_{xx}/\sigma_{0}+B^{\prime\prime}\left(  \sigma_{xx}/\sigma_{0}\right)
^{2}, \label{scaling-NLHE}%
\end{equation}
where $B=C+C_{1}+C_{11}$, $B^{\prime}=C_{0}-C_{1}+C_{01}+C_{10}-2C_{11}$ and
$B^{\prime\prime}=C_{00}+C_{11}-C_{01}-C_{10}$. Here all $C$'s are
zeroth-order homogeneous terms of scattering time \cite{Du2018}. In
particular, $C=\chi_{yxx}^{\text{bcd}}\rho_{xx}$, $C_{0\left(  1\right)
}=\chi_{yxx}^{\text{sj,ei(ep)}}\rho_{0\left(  1\right)  }$ and $C_{00\left(
11\right)  }=\chi_{yxx}^{\text{sjl,ei(ep)}}\rho_{0\left(  1\right)  }$.
Following the conventional paradigm of the anomalous Hall effect
\cite{Hou2015,Yue2017}, in the previous understanding Eq. (\ref{scaling-NLHE})
is viewed as a scaling relation when tuning temperature, and the $B$'s are
believed to be $T$ independent \cite{Du2018}.

According to our work on the anomalous Hall effect, however, it is apparent
that the $B$'s are in fact $T$ dependent. An experimental procedure similar to
the aforementioned one for the anomalous Hall effect can be applied to verify
this idea. The $T$ dependence of $C$ resulted from the Berry-curvature dipole
can be regarded to be weak \cite{Mak2019}, as is verified in the relaxation
time approximation \cite{Du2018}, under which $C$ is independent of scattering
time. Thus the $T$ dependence of the $B$'s arises from that of the
phonon-related side-jump and side-jump-like zeroth-order homogeneous terms,
e.g., $C_{1}$ and $C_{11}$. In the high-$T$ regime where $\rho_{1}\propto T$,
the $T$ dependence of electron-phonon scattering times in the denominator and
numerator of these terms are the same ($1/T$), and thereby drop out, leading
to $T$ independent $B$'s. While at lower temperatures, the bosonic
phonon-occupation leaves the $T$ dependence irreducible. Illustration of this
argument using a prototypical model of the nonlinear Hall effect, namely the
2D tilted massive Dirac model \cite{Fu2015,Konig2018,Du2018,Sodemann2019}, is
presented in the Supplemental Material \cite{Supp}.

In a recent experiment on the nonlinear Hall effect in bilayer WTe$_{\text{2}%
}$ \cite{Ma2019}, only the Berry-curvature dipole mechanism was claimed.
Whereas in another experiment on few-layer WTe$_{\text{2}}$ \cite{Mak2019}, a
scaling taking the form of (\ref{scaling-TYJ}) ($\sigma_{\text{AH}}\rightarrow
V_{y}^{N}/\left(  V_{x}^{L}\right)  ^{2}$) was observed in tuning-$T$
measurements, indicating the presence of the skew scattering. Since the
experimental system is dirty ($\eta<1$), the emergence of scaling
(\ref{scaling-TYJ}) in practice is reasonable. However, the tuning-$T$
measurement alone can not distinguish the side-jump (-like) contribution from
the Berry-curvature dipole. In order to further investigate the relevance of
the side-jump (-like) contribution, a possible route suggested by our results
is to observe the $T$ dependence of $B$ in Eq. (\ref{scaling-NLHE}) by using
multi-step WTe$_{\text{2}}$ samples through the above described experimental approach.

In summary, we have uncovered that, the conventionally identified scaling
parameters, which play the central role in the study of the anomalous Hall
effect, in fact depend on temperature below the equipartition regime. An
experimental approach has been proposed to observe this hitherto unexpected
temperature-dependence. We also showed that the similar physics applies to the
recently proposed scaling relations for the nonlinear Hall effect, and
provides a possible approach to identifying experimentally the relevance of the
side-jump contribution besides the Berry-curvature dipole.

\begin{acknowledgments}
We thank Ming Xie, Shengyuan A. Yang, Zongzheng Du, Haizhou Lu, Dazhi Hou and Tingxin Li for insightful discussions.
We are especially indebted to Zongzheng Du and Haizhou Lu for sharing their unpublished supplementary material.
Q.N. is supported by DOE (DE-FG03-02ER45958, Division of Materials Science and Engineering)
on the model analysis in this work. C.X. and H.Z. are supported by NSF (EFMA-1641101) and Welch Foundation (F-1255).
\end{acknowledgments}

\end{document}